\documentclass[aps,preprint,secnumarabic,nobalancelastpage]{revtex4}

\usepackage{graphics}      
\usepackage{graphicx}      
\usepackage{longtable}     
\usepackage{url}           
\usepackage{bm,color}            
\usepackage{amssymb, amsmath, enumerate, theorem, epsfig,subfigure,setspace}
\usepackage{amsfonts}

\begin{document}

\title{Global optimization of spin Hamiltonians with gain-dissipative systems}
 \author{Kirill P. Kalinin$^1$ and Natalia G. Berloff$^{2,1}$ }
\email[correspondence address: ]{N.G.Berloff@damtp.cam.ac.uk}
\affiliation{$^1$Department of Applied Mathematics and Theoretical Physics, University of Cambridge, Cambridge CB3 0WA, United Kingdom}
\affiliation{$^2$Skolkovo Institute of Science and Technology Novaya St., 100, Skolkovo 143025, Russian Federation}

\date{June 10, 2018}

\begin{abstract}{Recently, several  platforms were proposed and demonstrated a proof-of-principle for finding the global minimum of the spin Hamiltonians such as the Ising and XY models using gain-dissipative quantum and classical systems. The implementation of dynamical adjustment of the gain and coupling strengths has been established as  a vital feedback mechanism for analog Hamiltonian physical systems that aim to simulate spin Hamiltonians. Based on the principle of operation of such simulators we develop a novel class of  gain-dissipative algorithms for global optimisation of NP-hard problems and show its performance in comparison with the classical global optimisation algorithms. These systems can be used to study the ground state and statistical properties of spin systems  and as a direct benchmark for the performance testing of the gain-dissipative physical simulators.  The estimates of the time operation of the physical implementation of the gain-dissipative simulators for large matrices show a possible speed-up of the several orders of magnitude  in comparison with classical computations. }
\end{abstract}

\maketitle
Finding the global minimum of spin Hamiltonians has been instrumental in many areas of modern science.  Such Hamiltonians have initially been  introduced in condensed matter 
 to study
magnetic materials \cite{magnetic, ising} and by now they became fundamentally important in a vast spread of 
many other disciplines such as  quantum gravity
\cite{quantumgravity}, combinatorial optimization \cite{lucas}, neural networks \cite{nn}, protein structures \cite{protein}, error-correcting codes \cite{errorcorrection}, X-ray crystallography \cite{xcrystal}, diffraction imaging \cite{di}, astronomical imaging \cite{astro}, optics \cite{optics}, microscopy \cite{micro}, biomedical   applications \cite{biomed},
percolation clustering \cite{percolation} and machine learning \cite{chertkov}.

The  spin degrees of freedom in spin models are either discrete or continuous.  In particular, we will be concerned with the XY model, where spins lie on a unit circle $s_j= \cos \theta_j + {\rm i} \sin \theta_j$, the Ising model where spins take values $s_j=\pm 1$ and $q$-state Potts model where spins take $q$ discrete values. For $N$ spins the classical Hamiltonians for these models can be written as  
\begin{equation}
H=-\sum_{i=1}^N\sum_{j=1}^N J_{ij} \cos(\theta_i-\theta_j) + \sum_{i=1}^N g_i \cos\theta_i,
\label{h}
\end{equation}
where the elements $J_{ij}$ of matrix ${\pmb J}$ define the strength of the couplings between $i$-th and $j$-th spins represented by the phases $\theta_i$ and $\theta_j$, $g_i$ is the strength of the external field acting on spin $i$. For the continuous XY model $\theta_j\in [0,2 \pi)$, for the Ising model $\theta_j\in \{0,\pi\}$, and for the $q$-state Potts model $\theta_j=2\pi j/q, j=1,...,q$.

For a general matrix of coupling strengths ${\pmb J}$ finding the global minimum of such  problems is known to be strongly NP-hard \cite{ZHANG06} (with the decision problem to be NP-complete),  meaning that an efficient way of solving them can be used to solve all problems in the complexity class NP that includes a vast number of important problems such as partitioning, the travelling salesman problem, graph isomorphisms, factoring, nonlinear optimisation beyond quadratic, etc. For instance, the travelling salesman problem of a record size 85,900 has been solved by the state of the art Concorde algorithm in around 136 CPU-years \cite{Concorde2006}.  The actual time required to find the solution also depends on the matrix structure. For instance, for positive definite matrices, finding the global minimum of the XY model remains NP-hard due to the non-convex constraints but can be effectively approximated using an SDP relaxation \cite{candes11} with the  performance guarantee $\pi/4$  \citep{ZHANG06}. Sparsity also plays an important role: for sufficiently sparse matrices  fast methods exist \cite{gespar}.   As for many other hard optimisation problems, there are three types of algorithms  for solving  spin Hamiltonian  problems on a classical computer:  exact methods that find the optimal solution to the machine precision, approximate algorithms that generate the solution within a performance guarantee  and heuristic algorithms where suitability for solving a particular problem comes from some empirical testing \cite{empirical}. Exact methods can be used to solve small to medium matrix instances, as they typically involve  branch-and-bound algorithms and the exponential worst-case runtime.   The heuristic algorithms such as  simulated annealing can quickly deliver a decent, but suboptimal (and possibly infeasible) solution \cite{wang14}.   Finally, global minimization of the XY and Ising models is known to be  in APX-hard class of problems \cite{APX}, so there is no polynomial-time approximation algorithm that gives the value of the objective function that is arbitrarily close to the optimal solution (unless P = NP). The problem becomes even more challenging when the task is to find not only an approximation to  the global minimum of the objective function, but also the minimisers as needed for instance in  image reconstruction. The values of the objective functions can be very close, but for the entirely different sets of minimizers.

 Recently, several  platforms were proposed and demonstrated a proof-of-principle for finding the global minimum of the spin Hamiltonians such as the Ising and XY models using gain-dissipative quantum and classical systems: the injection-locked lasers, \cite{yamamoto11}, the network of optical parametric oscillators, \cite{yamamoto14,takeda18},  coupled lasers \cite{coupledlaser}, polariton condensates \cite{NatashaNatMat2017},  and photon condensates \cite{KlaersNatPhotonics2017}. In the gain-dissipative simulators the  phase of the so-called coherent centre (CC) is mapped into the ``spin" of the simulator. Such CC can be a condensate \cite{NatashaNatMat2017, KlaersNatPhotonics2017} or  a coherent state generated in a laser cavity \cite{takeda18,coupledlaser}. The underlying operational principle of such simulators depends on a gain process that is increased from below until a nonzero occupation appears via the supercritical Hopf bifurcation and the system becomes globally coherent throughout many CCs.  The coherence occurs at the maximum occupancy for the given gain.  It was suggested and experimentally verified  that the maximum occupancy of the system is related to the corresponding spin Hamiltonian \cite{NatashaNatMat2017}. When the heterogeneity in densities of the CCs is removed by dynamically adjusting the gain  the coherence will be established at the global state of the corresponding spin Hamiltonian \cite{GD-polariton}. We refer to these platforms as gain-dissipative analog Hamiltonian optimisers \cite{blockchain} that, in spite of  having  different quantum hardwares,  share the basic principle that suggests the convergence to the global minimum of the  spin Hamiltonian. 

 Here, motivated by the operation of such physical systems, we develop a new class of classical gain-dissipative algorithms for solving large-scale optimisation problems based on the Fokker-Plank-Langevin gain-dissipative equations written for a set of CCs. We show how the algorithm can be modified to cover various spin models: continuous and discrete alike. We demonstrate the robustness of such iterative algorithms and show that we can tune the parameters for the algorithm to work efficiently on various sizes and coupling structures.   We show that such algorithms can outperform  the standard  global  optimiser algorithms and have a potential to become the state of the art algorithm. Most importantly, these algorithms  can be used as a benchmark for the performance of the physical gain-dissipative simulators. Finally, this framework allows us to estimate the operational time for a physical realisation of such simulators to achieve the global minimum.  We show that for large problem sizes  the analog simulator when built would outperform the classical computer computations by several orders of magnitude.  

The paper is organised as follows. We formulate the general classical gain-dissipative algorithm for finding the global minimum of various spin Hamiltonians in Section 1. In Sections 2 and 3 we investigate its performance on global optimisations of the XY and Ising Hamiltonians  by comparing it to the state-of-the art global optimisers.  We conclude with the discussion of the performance of the actual physical systems in Section 4 and conclude with Section 5. %
\section{Gain-dissipative approach for minimising the spin  Hamiltonians}
The principle  of operation of the  gain-dissipative simulator with $N$ CCs for minimisation of the spin Hamiltonians given by Eq.~(\ref{h}) is described by the following set of the rate equations \cite{GD-polariton, GD-resonance}
\begin{eqnarray}
\frac{d\Psi_i}{d t} &=& \Psi_i  ( \gamma_i^{\rm inj} - \gamma_c - |\Psi_i|^2) + \sum_{j, j \neq i} \Delta_{ij} K_{ij} \Psi_j \nonumber \\
&+& \sum_{q=1}^n h_{qi} \Psi^{*(q-1)}_i + D \xi_i(t),
\label{main}
\end{eqnarray}
where $\Psi_i(t)$ is a classical complex function that describes the state of the $i$-th CC, $\gamma_i^{\rm inj}$ is the rate at which particles are injected non-resonantly  into the $i-$ state,  $\gamma_c$ is the rate of loosing the particles, the coupling strengths are represented by $\Delta_{ij} K_{ij}$ where we separated the effect of the particle injection that changes the strength of coupling represented by  $\Delta_{ij}$ from the other coupling mechanisms   represented by $K_{ij}$. We consider two cases $\Delta_{ij}=1$ that physically corresponds to the site dependent  dissipative coupling and $\Delta_{ij}=\gamma_i^{\rm inj}(t) +  \gamma_j^{\rm inj}(t)$ appropriate for the description of the geometrically coupled condensates \cite{GD-polariton}. We also include the complex function $\xi_i(t)$ that represents the white noise with a diffusion coefficient $D$ which disappears at the threshold. The coefficients $h_{qi}$ represent the strength of the external  field  with the resonance $q:1$ \cite{GD-resonance}. Compared to the actual physical description \cite{GD-polariton, GD-resonance}, in writing Eq.~(\ref{main}) we neglected  the possible self-interactions within the CC and re-scaled $\Psi_i$ so that the coefficient at the nonlinear term $|\Psi_i|^2 \Psi_i$  is $1$ and allowed for several ($n$) resonant terms to be included. By writing $\Psi_i=\sqrt{\rho_i}\exp[{\rm i} \theta_i]$ and separating real and imaginary parts in Eq. (\ref{main}) we get the equations on the time evolution of the number density $\rho_i$ and the phase $\theta_i$
\begin{eqnarray}
\frac{1}{2}\dot{\rho}_i(t)&=&(\gamma_i^{\rm inj} - \gamma_c- \rho_i) \rho_i 
+ \sum_{j;j\ne i} \Delta_{ij}^{\rm inj}K_{ij} {\sqrt{\rho_i\rho_j}}\cos\theta_{ij}\nonumber \\ &+& \sum_{q=1}^nh_{qi} \rho_i^{\frac{q}{2}}\cos (q \theta_i),\label{rho}\\
\dot{\theta}_i(t)&=&-\sum_{j;j\ne i} \Delta_{ij}^{\rm inj}K_{ij} {\frac{\sqrt{\rho_j}}{\sqrt{\rho_i}}} \sin\theta_{ij} \nonumber\\
&-& \sum_{q=1}^nh_{qi}\rho_i^{\frac{q}{2}-1}\sin(q \theta_i),\label{theta}
\end{eqnarray}
where $\theta_{ij}=\theta_i-\theta_j$. 

As we have previously shown \cite{GD-polariton, GD-resonance}, the individual control of the pumping rates $\gamma_i^{\rm inj}$ is required to guarantee that the fixed points of the system coincide with minima of the spin Hamiltonian given by Eq.~(\ref{h}).  As the injection rates $\gamma_i^{\rm inj}$ raise from zero they have to be adjusted in time to bring all CCs to condense at the same specified number density $\rho_{\rm th}$. Mathematically, this is achieved by 
  \begin{equation}
 \frac{d \gamma_i^{\rm inj}}{dt} = \epsilon (\rho_{\rm th} - \rho_i),
 \label{gamma}
 \end{equation}
 where $\epsilon$ controls the speed of the gain adjustments. If we take $\Delta_{ij}=1$ we assign $K_{ij}=J_{ij}$. If $\Delta_{ij}$ depends on the injection rates, the coupling strengths will be modified, so they have to be adjusted as well to bring the required coupling $J_{ij}$ at the  fixed point by
 \begin{equation}
\frac{d K_{ij}^{\rm inj}}{dt} = \hat\epsilon (J_{ij} - \Delta_{ij} K_{ij}),
 \label{gamma2}
 \end{equation}
where $\hat{\epsilon}$ controls the rate of the coupling strengths adjustments. 
Equation (\ref{gamma2}) indicates that the couplings need to be reconfigured depending on the injection rate: if the coupling strength scaled by the gain at time $t$ is lower (higher) than the objective coupling $J_{ij}$, it has to be increased (decreased) at the next iteration. We shall refer to  numerical realisation of Eqs.~(\ref{main}, \ref{gamma}) and  (\ref{main}, \ref{gamma}-\ref{gamma2}) as the  `GD algorithm'  and the `GD-mod algorithm' respectively. 
 
The fixed point of Eqs. (\ref{rho}-\ref{gamma2}) are
 \begin{eqnarray}
 \rho_i&=&\rho_{\rm th}=\gamma_i^{\rm inj} -\gamma_c +  \sum_{j;j\ne i} J_{ij} \cos\theta_{ij}\nonumber\\
 &+&\sum_q h_{qi} \rho_{\rm th}^{\frac{q}{2}-1} \cos(q \theta_i),
 \label{pth}
 \end{eqnarray}
with the total number of particles in  the system given by $M=N\rho_{\rm th}=\sum_i\gamma_i^{\rm inj} - N\gamma_c +  \sum_{i,j;j\ne i} J_{ij} \cos\theta_{ij} + \sum_q \rho _{\rm th}^{\frac{q}{2}-1} \sum_i h_{qi} \cos(q \theta_i).$ Such a value of the total number of particles will be first reached at the minimum of  $\sum_i\gamma_i^{\rm inj}$, therefore, at the  minimum of the spin  Hamiltonian given by
\begin{equation}
H_s = -\sum_{i,j;j\ne i}  J_{ij}\cos\theta_{ij} - \sum_q \rho _{\rm th}^{\frac{q}{2}-1} \sum_i h_{qi} \cos(q \theta_i).
\label{hs}
\end{equation}
 Eq.~(\ref{hs}) represents the general functional that our GD and GD-mod algorithms optimise. By choosing which $h_{qi}$ are non-zero we can emulate a variety of spin Hamiltonians. If $h_{qi}=0$, then Eq.~(\ref{hs}) represents the XY Hamiltonian. If only $h_{2i}=h_2$ are non-zero with $h_2> \sum_{j;j \ne i} |J_{ij}|$ for any $i$, then the second term of the right-hand side of the Hamiltonian  (\ref{hs}) represents the penalty forcing phases to be $0$ or $\pi$ which, therefore, leads to the Ising Hamiltonian. If only  $h_{qi}=h_q$ for $q>2$ are non-zero, then Eq.~(\ref{hs}) emulates the $q$-state Potts model with phases restricted to discrete values $\theta_i=2\pi i/q$. Finally, introducing non-zero $h_{1i}$ together with non-zero $h_q$ for $q>1$ brings the effect of an external field of strength $g_i=h_{1i}/\sqrt{\rho_{\rm th}}$ in agreement with Eq.~(\ref{h}).
 
 \section{Global minimization of the XY Hamiltonian: $h_{qi}=0.$ }
 To find the global minimum of the XY Hamiltonian  we numerically evolve  Eqs.~(\ref{main},\ref{gamma}) with $h_{qi}=0$  using the 4th-order Runge-Kutta integration scheme.
 
\begin{figure}[t!]
\centering
  \includegraphics[width=8.6cm]{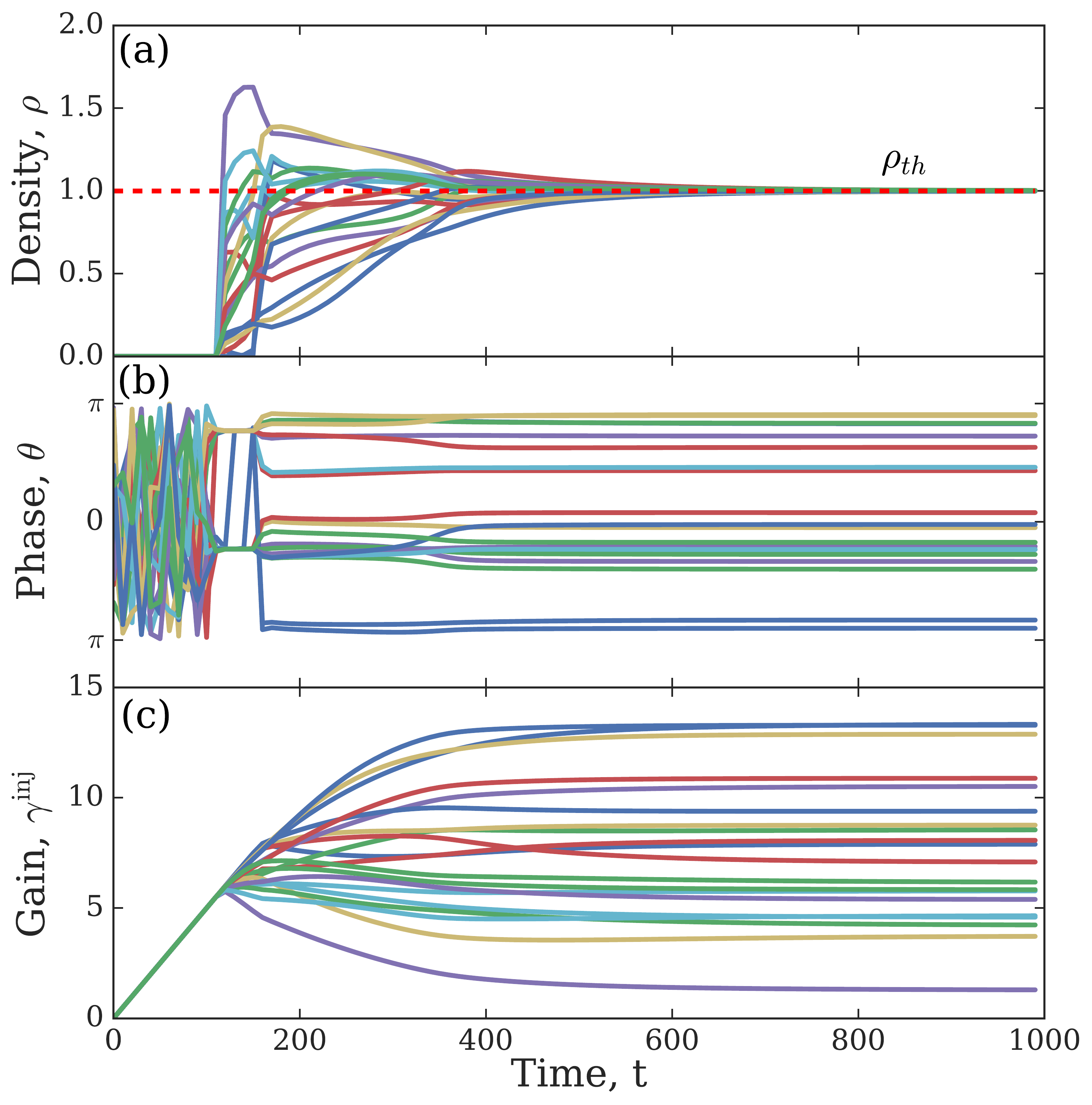}
\caption{Plots of (a) the number densities $\rho_i = |\Psi_i|^2$ of CCs, (b) the phases $\theta_i$ and (c) the injection rates $\gamma_i^{\rm inj}$ as functions of time obtained by the numerical integration of Eqs.~(\ref{main},\ref{gamma}) with $h_{ni}=0$ starting with zero initial conditions for $i=1,...,20$. }
 \label{Figure1}
\end{figure}

To illustrate    the operational principle of GD algorithm  for  minimising the XY Hamiltonian we consider  $N=20$ nodes and the coupling strengths $J_{ij}$ that are randomly distributed between $-10$ and $10$, see Fig.~\ref{Figure1}.  Starting from a zero initial condition $\Psi_i=0$, at the first stage of the evolution (while dimensionless $t<120$) the densities are well below the threshold (Fig.~\ref{Figure1}a), phases span various configurations  (Fig.~\ref{Figure1}b), and  all injection rates are the same (Fig.~\ref{Figure1}c).  Then the nodes  start reaching and in some cases overcoming  the threshold, the injection rates start being individually adjusted to bring all the nodes to the same value while phases stabilise to realise the minimum of the XY Hamiltonian. 

A numerical approach for solving NP-hard optimisation problems depends on the scale of the problem: intermediate-scale problems can be solved with general programming tools while large-scale problems require sophisticated algorithms that exploit the structure of a particular type of objective function and are usually solved by iterative algorithms. Since the proposed GD method based on Eqs.~(\ref{main},\ref{gamma}) is an iterative algorithm, we aim to investigate its two main aspects. First, we conduct the global convergence analysis on small and mid-scale problems and verify that the algorithm converges to a global minimum. The fact that the minimum is truly global we confirm by exploiting other optimisation methods. As for any heuristic iterative algorithm, such convergence properties can be established with confidence by performing numerous numerical experiments on different problems. Second, we perform the complexity analysis on large-scale problems with a focus on how fast the algorithm converges showing, in particular, that the GD algorithm gives  a polynomial-time growth per iteration in the cases  of general dense matrices.

To characterise the performance of the GD algorithm, we compared it to the heuristic global optimisation solvers such as direct Monte Carlo sampling (MC)  and the basin-hopping (BH) algorithm. Both methods depend on a local minimisation algorithm for the optimal decent to a local minimum at each iteration. We considered several local minimisation methods as applied to the minimization of the XY Hamiltonians   and determined that 
the quasi-Newton method of Broyden, Fletcher, Goldfarb, and Shanno (L-BFGS-B) \citep{BFGS_1995,BFGS_1997} has shown the best performance (see Appendix). The L-BFGS-B algorithm is a local minimisation solver which is designed for large-scale problems and shows a good performance even for non-smooth optimisation problems \citep{BFGS_1995,BFGS_1997}. At each iteration of the MC algorithm, we generate a random starting point and use L-BFGS-B algorithm to find the nearest local minimum.  These minima are compared  to find the global minimum. The BH algorithm is a global minimisation method that has been shown to be extremely efficient for a wide variety of problems in physics and chemistry \citep{BasinHopping1997} and to give a better performance on the spin Hamiltonian optimisation problems  than other heuristic methods such as simulated annealing \cite{simAnneal83}. It is an iterative stochastic algorithm that at each iteration uses  a random perturbation of the coordinates with a local minimisation followed by the acceptance test of new coordinates based on the Metropolis criterion.  Again L-BFGS-B algorithm  has shown the best performance as a local optimiser at each step of the BH  algorithm. Both BH and MC algorithms were supplied with the analytical Jacobian  of the objective function for  better performance results.

%
%

%
\begin{figure}[b!]
\centering
  \includegraphics[width=8.6cm]{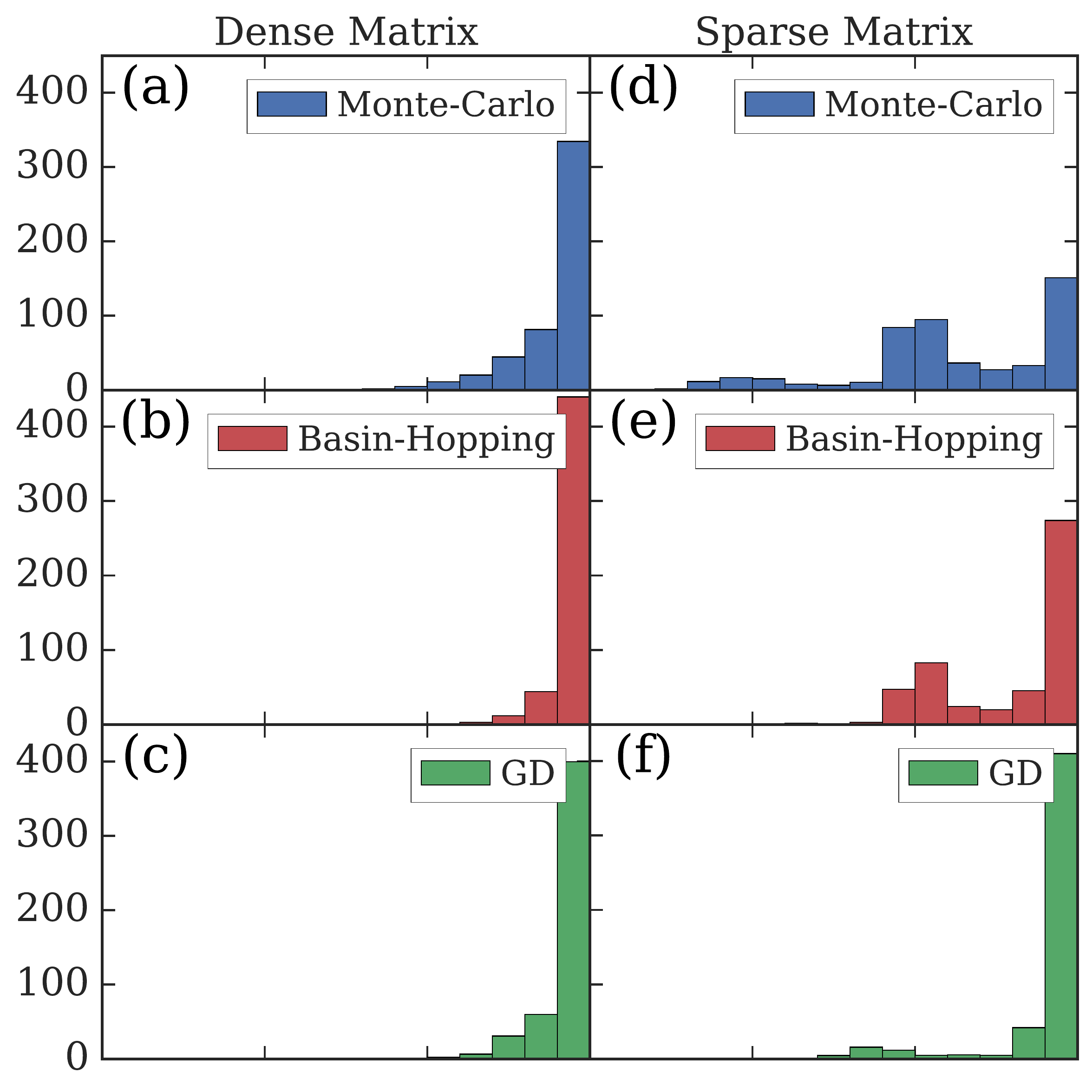}
\caption{The success probability of (a,d) MC, (b,e) BH, and (c,f) GD  algorithms when minimising the XY Hamiltonian for the matrix size $N = 50$.  The results of 500 runs are averaged over 50 real symmetric coupling matrices ${\bf J}$ with the elements randomly distributed  in  $[-10,10]$ for (a-c) `dense' and (d-f) `sparse' matrices described in the main text. The number of internal BH iterations was set to ten to bring about a similar performance to the GD algorithm for `dense' matrices.}
 \label{Figure3}
\end{figure}

To confirm the global convergence, we compared the GD algorithms to BH and MC algorithms by  minimizing XY Hamiltonian for various matrices using the parameters listed in \cite{numer}. In particular, we generated 50  real symmetric coupling matrices  ${\bf J}=\{J_{ij}\}$ of two types. We considered `dense' matrices  with elements that are randomly distributed in  $[-10, 10]$ and `sparse' matrices where  each CC is randomly connected to exactly three other CCs with the coupling strengths  randomly generated from the interval with the bounds that are randomly taken from $\{-10,-3,3,10\}$. For each such matrix, we ran the GD, BH and MC algorithms starting from 500 random initial conditions for BH and MC algorithms and from zero initial conditions and 500 different noise seeds for the GD algorithm. The values of the global minimum of the objective function found by GD algorithm and the comparison methods were found to match to ten significant digits.
For `dense' matrices the success probabilities of the GD algorithms were similar to both comparison methods. The distribution of success probabilities over the various `dense' matrix instances is shown in Fig.~\ref{Figure3}(a-c) for  $N = 50$ and suggests that for such matrices the systems have very narrow spectral gap so the distributions are densely packed for probabilities over $93\%$ for the  MC, $96\%$ for the BH, and $95\%$ for the GD algorithm.  However, the GD algorithm greatly outperforms the comparison algorithms on `sparse' matrices as Figs.~\ref{Figure3}(d,e,f) illustrate. The structure  of such matrices makes the barriers between local and global minima higher, and, therefore, worsens the  performance of BH and MC methods, but barely has an  affect on the GD algorithm that approaches the global minimum from below. Thus, we established the global convergence properties of the proposed GD algorithms on various problems and verified that the GD algorithms converge to the global minimum. The further advantages of the GD algorithms over the best classical optimisers for some special types of  the coupling matrices are  elucidated elsewhere \cite{kirillFuture}.

 \section{Global minimization of the Ising Hamiltonian: $h_{2i}=h_2\ne0.$}
\begin{figure}[t!]
\centering
  \includegraphics[width=8.6cm]{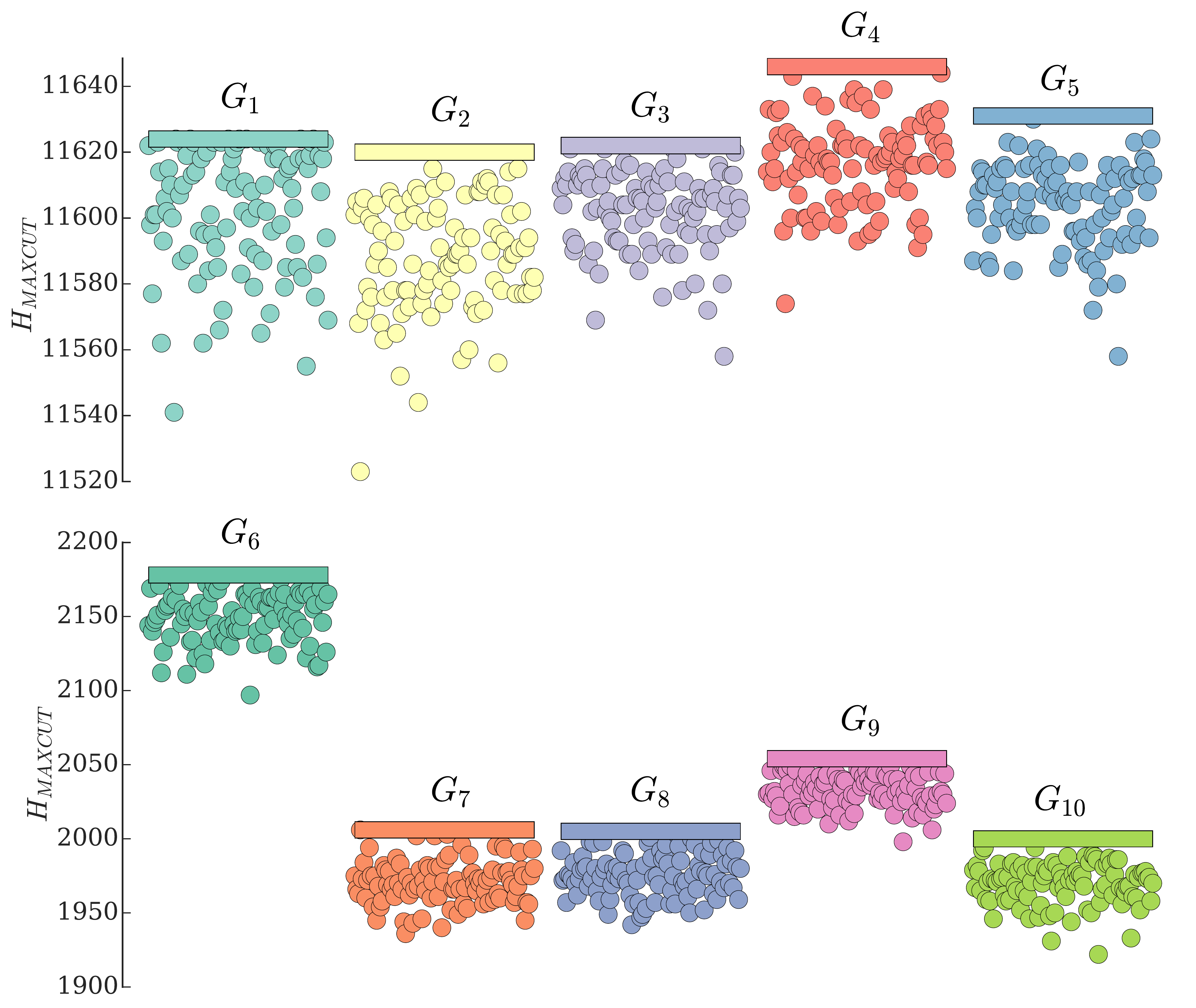}
\caption{The performance of the GD algorithm (\ref{main}-\ref{gamma}) for solving the Max-Cut optimisation problem on G-Sets $\{G_1-G_{10}\}$ of size $N = 800$. The  known optimal values are plotted with coloured rectangles for each $G_i$.  The results of the GD algorithm are shown with scatters for 100 runs on each $G_i$. All found  Max-Cut values are within $0.8\%$ ($4.5\%$) deviation from the optimal solution  for $G_1-G_5$  ($G_6-G_{10}$) sets. The time per each run of the GD algorithm has been fixed to around $35-40sec$ for all G-Sets.}
 \label{Figure4}
\end{figure}

To find the global minimum of the Ising Hamiltonian  we solve Eqs.~(\ref{main},\ref{gamma}) with $h_{qi}=0$ if $q\ne 2$ and $h_{qi}=h_2$ numerically.
Based on these equations we test  the GD algorithm  by finding the minima of Max-Cut optimisation problem on the well-known G-Set instances \cite{Gsets} and summarise our findings in Fig.~\ref{Figure4}. The optimal Max-Cut values \cite{BLS2013} are plotted with coloured rectangles and the solutions of the GD algorithm are shown with scatters for 100 runs for each G instance. The algorithm demonstrates a good performance in terms of solution quality with the average found cuts being within $0.2-0.3\%$ for $G_1-G_5$ and $1.1-1.8\%$ for $G_6-G_{10}$ from  the optimal solutions.  The same   same numerical parameters were used for all simulations \cite{numer} and  the computational time for  finding each cut has been limited by the same value ($35-40sec$ \cite{laptop}) for all G-Sets. The time performance of the state-of-the art algorithms is highly dependent on a particular problem and for $G_1-G_{10}$ varies from $13sec$ to $317sec$ \cite{BLS2013laptop} for \textit{breakout local search} algorithms \cite{BLS2013} and is within $100-854sec$ for \textit{GRASP tabu search} \cite{TabuSearch2013}, though their solutions are much less deviated from the optimal values. Therefore, the proposed GD algorithm is highly competitive with the existing state-of-the art Max-Cut algorithms  at least in terms of the computational time. The deviation of solutions from the optimal values can be further reduced by tuning the  parameters $\rho_{th}$ and $\epsilon$ or by investigating the extensions to the suggested GD algorithm. Among such possible add-ons to the GD-algorithm can be the introduction of individual dynamic rates of the gain adjustments $\epsilon_i(t)$.

\section{Projected performance of the GD simulators}
So far we discussed the implementation of the GD algorithms on a classical computer. An actual physical implementation of these algorithms on simulators will enjoy a super-fast operation and parallelism in processing various phase configurations as the system approaches the global minimum from below even if the system behaves fully classically. Further acceleration could be expected if quantum fluctuations and quantum superpositions  contribute to processing of the phase configurations. The times involved into the hardware operation  of the GD simulators  vary on the scale of pico- to milli-seconds. For instance, in the system of non-degenerate
optical parametric oscillators (NOPO) the time-devision multiplexing method is used to connect a large number of nodes and  the couplings are realised  by  mutually injecting with optical delay lines with the cavity round trip time being of the order of $\mu$s \cite{takeda18}, it takes an order of  100  picoseconds for the polariton graphs  to condense \cite{NatashaNatMat2017} and 10 ps to 1 ns for photon condensates \cite{KlaersNatPhotonics2017}. The feedback mechanism can be implemented via optical delay lines (in NOPO system), by holographic reconfiguration of  the injection via the spatial light modulator or mirror light masks (e.g. by DLP high-speed spatial light modulators) in solid-state condensates or  by electrical injection (e.g. in the polariton lattices \cite{electrical18}). 

\begin{figure}[t!]
\centering
  \includegraphics[width=8.6cm]{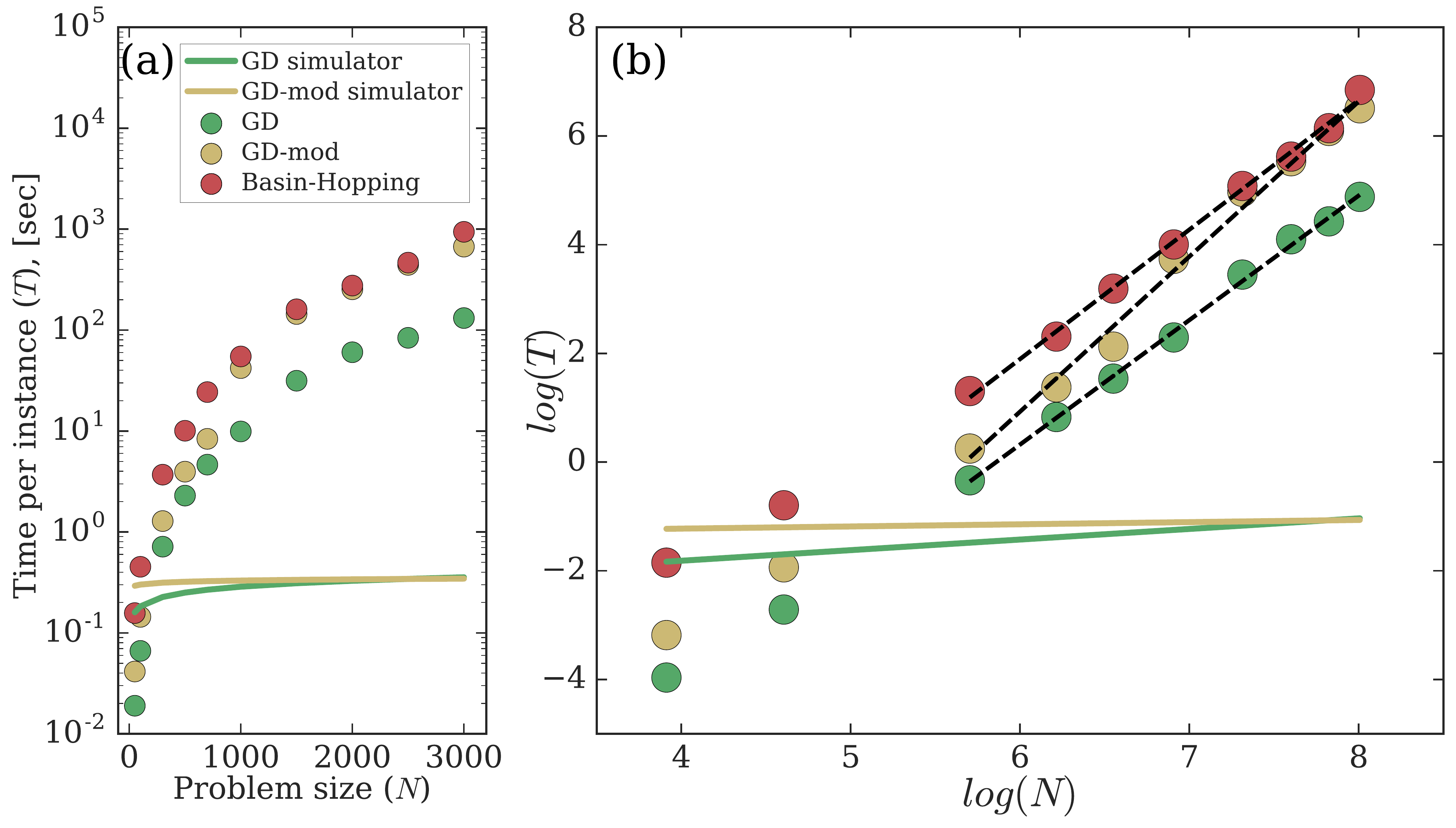}
\caption{ The performance of the GD, GD-mod, and BH algorithms in minimizing the XY Hamiltonian with  $N$ up to 3000. (a) The time per instance $T$ as a function of the problem size $N$. In the case of GD and GD-mod algorithms, $T$ is  the time averaged over 20 runs necessary to reach a stationary state. For the BH algorithm, this time is per ten internal BH iterations necessary to have about the same success probabilities as the GD algorithm. (b) $T$ as a function of $N$ in the logarithmic scale. The performance of the algorithms are fitted by the linear interpolation functions $2.29 \log N - 13.42$, $2.85\log N - 16.2$, and $2.38\log N - 12.39$, for the GD, GD-mod, and BH algorithms, respectively. The projected performance of the GD simulator dominated by the dissipative (gain) coupling is shown with solid green (yellow) lines whose linear asymptotic in (b) is $0.2 \log N - 2.6$ ($0.04 \log N - 1.38$).}
 \label{Figure5}
\end{figure}

The number of iterations one needs to reliably find the global minimum grows with the size of the problem $N$.  This growth is expected to be exponential for any algorithm (if $P\ne NP$). However, we can compare how time per iteration grows with the problem size for considered algorithms. 
We perform the complexity analysis per iteration on mid- and large-scale problems and summarise the results in Fig.~\ref{Figure5}. The GD algorithm demonstrates the consistent speedup over BH algorithm for all problem sizes $N$ in Fig.~\ref{Figure5}(a). The log plot in Fig.~\ref{Figure5}(b) indicates that both algorithms show polynomial time per iteration with the complexity of the GD algorithm being close to ${\cal O}(N^{2.29})$. 

As the feedback mechanism has to be implemented we need to factor in the time it takes and the necessary number of such adjustments which corresponds to the number of internal iterations for the GD and GD-mod algorithm per each run.  By taking an upper limit on the feedback time as $0.1ms$ and using an average number of iterations of the GD and GD-mod algorithms we can estimate the upper bound on the time needed by the physical implementation of the GD simulator to find the global minimum. In Fig.~\ref{Figure5} we show  such estimates by the solid green (yellow) lines for the GD (GD-mod) simulators. For large $N$ from Fig.~\ref{Figure5} we estimate the speed-up of the GD simulators in comparison with the classical computations to be of the order  of $10^{-5}N^{2} - 10^{-7} N^3$. For $N$ of the order of ten thousands this gives the speed-up of at least four orders of magnitude. Because of  the adaptive setting of the coupling matrix in the GD-mod algorithm, the number of internal iterations  grows  slower with the size of problem $N$ than for the GD algorithm so that the performance of the GD simulator  can  possibly be surpassed by the GD-mod simulator  for large $N$.
\section{Conclusions} 
Motivated by a recent emergence of a new type of analog Hamiltonian optimisers -- the gain-dissipative simulators -- we formulate a novel gain-dissipative algorithm for solving large-scale optimisation problems which is easily parallelisable and can be efficiently simulated on classical computers. We show its computational advantages in comparison with the state-of-the-art methods. We argue that the generalisation of the GD algorithm for solving different classes of $NP$-hard problems can be done for both continuous and discrete problems and demonstrate it by solving  quadratic continuous and binary optimisation problems. The GD algorithm  has a potential of becoming a new optimisation algorithm superior to other global optimisers. This algorithm allows us to formulate the requirement for the simulators hardware built using  a system of gain-dissipative oscillators of different nature. Our algorithm, therefore, can be used to benchmark the existing gain-dissipative simulators.  When the run-time of the classical algorithm is interpreted in terms of the time of the actual operation of the physical system one would expect such simulators to greatly outperform the classical computer.


Finally, we would like to comment on classical vs quantum operation of such simulators. When a condensate (a coherent state) is formed -- the system behaves classically as many bosons are in the same
single-particle mode and non-commutativity of the field operators can be neglected.  However, the condensation process by which the global minimum of the spin models is found involves quantum effects. It was shown before, that the condensation process can be described by a fully classical evolution of the Nonlinear Schr\"odinger equation  that takes into account only stimulated scattering effects and neglects spontaneous scattering \cite{berloffSvistunov}. The classical or
quantum assignment to gain-dissipative simulators depends on whether 
quantum fluctuations and spontaneous scattering effects during the condensation provide a speed-up in comparison with fully classical noise and stimulated scattering.  This is an important question to address in the future research on such simulators and the comparison with the classical algorithm that we developed based on the gain-dissipative simulators architecture allows one to see if the time to find the solution scales better than with the
best classical algorithms.

\section*{Acknowledgements}
The authors acknowledge financial support from the NGP MIT-Skoltech. K.P.K. acknowledges the financial support from Cambridge Trust and EPSRC.
\section*{Appendix: Performance of local optimisation algorithms for the global minimization of the XY Hamiltonian} 

\begin{figure}[t!]
\centering
  \includegraphics[width=8.6cm]{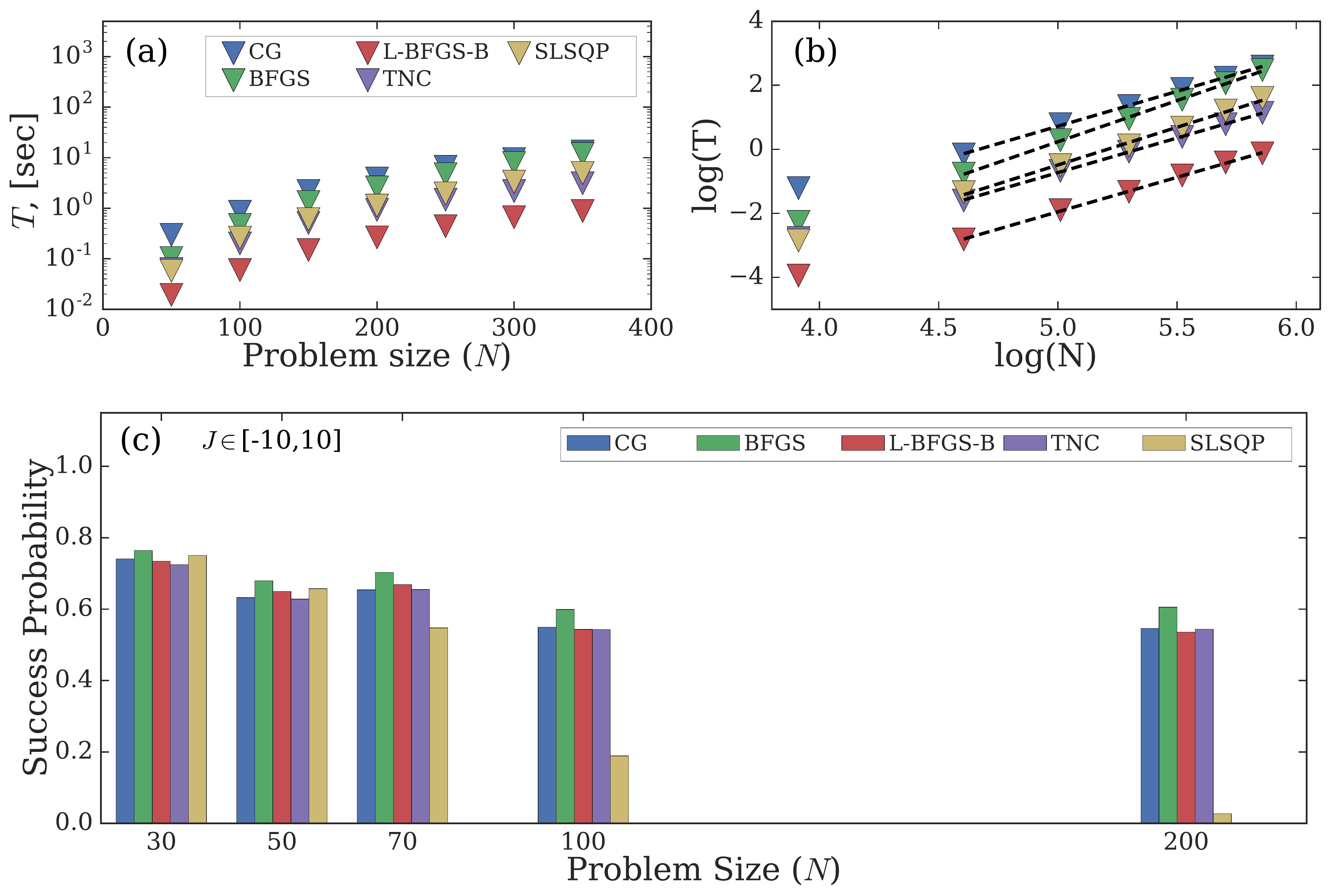}
\caption{The performance of the various local optimisers when finding the global minimum of the XY Hamiltonians for matrix sizes  up to $N = 200$. The run-time dependence on  the matrix size $N$ is shown in (a) and in a log scale in (b). The success probability of $99\%$ is shown in (c) where each algorithm starts from the same 100 random initial states. The probabilities are averaged over 25 dense coupling matrices with randomly generated elements in $[-10,10]$.}
 \label{FigureN}
\end{figure}

At  each iteration of the Monte-Carlo and the Basin-Hopping methods we use the  L-BFGS-B algorithm, since it has shown better results for  the global minimization of the XY Hamiltonian in terms of both performance and the quality of solution compared to other available algorithms in scipy.optimize.minimize library such as the  sequential least squares programming (SLSQP), nonlinear conjugate gradient algorithm (CG), truncated Newton (TNC) algorithm, and BFGS. In comparison with  the BFGS algorithm, the L-BFGS-B (limited memory BFGS) algorithm exploits an estimation of the inverse Hessian matrix. Each algorithm was  supplied with the analytical Jacobian. The performance of the algorithms is shown in Fig.~\ref{FigureN}. The L-BFGS-B algorithms is the fastest in comparison with all the  other algorithms  (see Fig.~\ref{FigureN}(a,b)) while the success probabilities are comparable (Fig.~\ref{FigureN}(c)). 


\begin{thebibliography}{00}
\bibitem{numer} The numerical parameters and the initial conditions for GD algorithm Eqs.~(\ref{main},\ref{gamma}) are as follows: $\epsilon =  \epsilon_0  \max_i \sum_j |J_{ij}|$,  $\rho_{th}= \rho_{0, th} \max_i \sum_j |J_{ij}|$, $U = 0$, $\sigma=1$, $\gamma (t=0)= - \max_i \sum_j |J_{ij}|$, $\rho_i (t=0)= 0$, $\theta_i (t=0)= 0$. For  GD-mod algorithm Eqs.~(\ref{main},\ref{gamma},\ref{gamma2}) we choose $\hat{\epsilon} = 1$ in addition to the parameters listed above. For the XY problem $\epsilon_0 = 0.05$ and $\rho_{0, th} = 0.1$, for the Max-Cut $\epsilon_0 = 0.005$ and $\rho_{0, th} = 0.15$. For the Max-Cut problems an external resonant field was slowly activated with $h_2 = 0.05(\tanh(3t/t_{max} - 1.5) + 1) \cdot \max_i \sum_j |J_{ij}|$.
\end{thebibliography}
\end{document}